\documentstyle[aps,pra,psfig,twocolumn]{revtex}

\begin{document}
\draft

\title{Discrimination of the Bell states of qudits by means
       of linear optics}

\author{Miloslav Du\v{s}ek}

\address{Institute of Physics, Slovak Academy of Sciences,
     D\'{u}bravsk\'{a} cesta 9, 842\,28 Bratislava, Slovakia}

\address{Department of Optics, Palack\'y University,
     17.~listopadu 50, 772\,00 Olomouc, Czech~Republic}

\date{\today}

\maketitle

\begin{abstract}
The question of the discrimination of the Bell states of
two qudits (i.e., $d$-dimensional quantum systems) by means of
passive linear optical elements and conditional measurements is
discussed. A qudit is supposed to be represented by $d$ optical
modes containing exactly one photon altogether.
From recent results of Calsamiglia it follows that
there is no way how to distinguish the Bell states of two qudits for
$d>2$ --- not even with the probability of success lower than one ---
without any auxiliary photons in ancillary modes.
Following the results of Carollo and Palma it is proved that it is
impossible to distinguish even only one such a Bell state with
certainty (i.e., with the probability of success equal to one),
irrespective of how many auxiliary photons are involved.
However, it is shown that auxiliary photons can
help to discriminate the Bell states of qudits with the high
probability of success: A Bell-state analyzer based on the idea of
linear optics quantum computation that can achieve the probability of
success arbitrarily close to one is described. It requires many
auxiliary photons that must be first ``combined'' into entangled
states. \end{abstract}




\section{Introduction}

Quantum optics and particularly linear optical elements
represent important tools for the experimental investigation of
the basic features of quantum information transfer and processing and
fundamentals of quantum theory. The special interest is
devoted to the study of entangled states that are not only highly
interesting by themselves but that find also the use in quantum
teleportation, quantum dense coding, quantum cryptography, quantum
computing, etc. Closely related to many of these issues is the so
called Bell-state analysis, i.e., the discrimination of maximally
entangled states completing the nonlocal orthonormal basis of two- or
multi-partite quantum system.

The relative success concerning the discrimination of two of the four
Bell states of two qubits with certainty by passive linear optical
elements \cite{Innsbruck,BellAn2} challenges the question whether it
is possible to do something similar also for the Bell states of two
qudits if their dimension $d>2$. At the first glance the direct
generalization of the approach given in Ref.\ \cite{BellAn2}
indicates that the limitations on the probability of success could be
less restrictive for $d>2$ than for $d=2$. But when one starts to
play with possible extensions of the original Innsbruck scheme
\cite{Innsbruck} for qudits he quickly gets into troubles. The
recent results of Calsamiglia \cite{calsamig} indicate that it is
impossible without additional auxiliary photons. This was the
motivation to subject the problem of ``generalized'' linear optics
Bell-state measurement to more detailed analysis.

In the considered  scheme, qudits are represented by $d$ modes of
radiation (with the same frequencies and polarizations). The total
number of photons in these $d$ modes is required to be one. The
$i$-th ``logical'' (or computational) basis state corresponds to the
situation when exactly one photon is present in the $i$-th mode. Such
an implementation allows us to realize any unitary operation on a
single qudit (up to a global phase) in a deterministic way by the
means of passive linear optical elements \cite{Reck}.

In fact, we restrict our tools to beam splitters, phase shifters and
delay lines. All of them may be electronically switched --- the
conditionally dynamics is allowed, i.e., some operations may depend
on the result of the measurement on selected modes outgoing the
``previous'' operation. We consider ideal detectors that can
distinguish the number of impinging photons (in a given mode). Even
if such detectors are not available in practice yet our choice is
justified as we seek for fundamental limitations of linear optical
devices.

By the Bell states of two qudits we mean the maximally entangled
states of the following form (see, e.g., Ref.\ \cite{teleport})
\begin{equation}
  |\psi_{mn}\rangle = \frac{1}{\sqrt{d}} \sum_{j=0}^{d-1}
  \exp \left[ 2 \pi i \frac{j n}{d} \right]
  {\sf a}^{\dag}_{j} {\sf b}^{\dag}_{j \oplus m}
  |\mbox{vac}\rangle,
\label{Bell}
\end{equation}
where ${\sf a}^{\dag}_{k}$ and ${\sf b}^{\dag}_{k}$ are
bosonic creation operators in corresponding modes of the first and
the second qudit, respectively, $|\mbox{vac}\rangle$ denotes a vacuum
state, and $j \oplus m = (j+m) \mbox{~mod~} d$; $m$ and $n$ go from 0
to $d-1$.


\section{No-go theorem for the case without auxiliary photons}
\label{no-go}

Calsamiglia has shown \cite{calsamig} that any linear optical device
that does not use auxiliary photons in ancillary modes (however, that
may include conditional dynamics) cannot unambiguously discriminate
any state with Schmidt rank {\em higher than two} from any set of
two-qudit states spanning the whole two-qudit Hilbert space (qudits
are supposed to be represented by one-photon states over $d$ modes).
This is true even if the probability of successful discrimination is
allowed to be lower than one. If there is a nonzero probability of
some detection event for some state from the set that has
Schmidt rank higher than two then there is always at least
one another state from the set that gives a nonzero probability for
this detection event too.

It directly follows that there is no way how to distinguish
any one of $d^2$ Bell states of two qudits (for $d>2$) without error
by means of linear optics if no auxiliary photons are involved.


\section{Impossibility to discriminate a Bell-state with certainty}
\label{photons}

In the paper of Carollo and Palma \cite{discrimin} it is shown
under the same conditions as assumed here (i.e., linear
elements, arbitrary number of auxiliary modes, conditional
measurements, and photon number detectors are assumed) that any
two $L$-photon states over $M$ modes, randomly chosen from a known
set of $K$ states, are completely (i.e., with certainty)
distinguishable in the presence of auxiliary photons only if they are
completely distinguishable in the absence of auxiliary photons. It
also means that if some two states are not completely distinguishable
in the case when no auxiliary photons are involved, than they cannot
be distinguished even if any finite number of auxiliary photons are
employed.

As stated earlier it is not possible in any way to distinguish any
one of $d^2$ Bell states from all of the remaining states by linear
optical device with no additional photons if $d>2$. In other words,
to each Bell state there is at least one another Bell state such that
these two states cannot be distinguished from each other (providing
$d>2$). According to the statement given in the previous paragraph
this must stay valid even if auxiliary photons are allowed. So it is
impossible to discriminate even one of $d^2$ Bell states with
certainty (with 100\,\% probability of success) irrespective whether
auxiliary photons are allowed or not.


\section{Efficient Bell-state analyzer with linear optics}
\label{loqc}

Now we will show how to distinguish all the Bell states of two qudits
with the probability of success arbitrarily close to one using only
linear optical elements (and auxiliary photons).

Recently, a scheme for ``non-deterministic'' quantum computation with
linear optics were proposed by Knill, Laflamme, and Milburn
\cite{LOQC_nature,LOQC_qph}. It is based on a non-linear phase shift:
$$
\alpha_0 |0\rangle + \alpha_1 |1\rangle + \alpha_2 |2\rangle \to
 \alpha_0 |0\rangle + \alpha_1 |1\rangle - \alpha_2 |2\rangle,
$$
where kets represent number states in a given mode. We will denote
this operation ``NS''. A simple device was designed that can perform
this operation with probability 1/4. It consists of a couple of beam
splitters and phase shifters. The effective non-linearity is provided
by a measurement process. Two ancillary modes, with a single photon
in one of them, and two photon-number detectors are necessary. The
successful operations turns up when the first detector registers one
and the second detector no photon.

\begin{figure}
 \centerline{\psfig{width=0.7\hsize,file=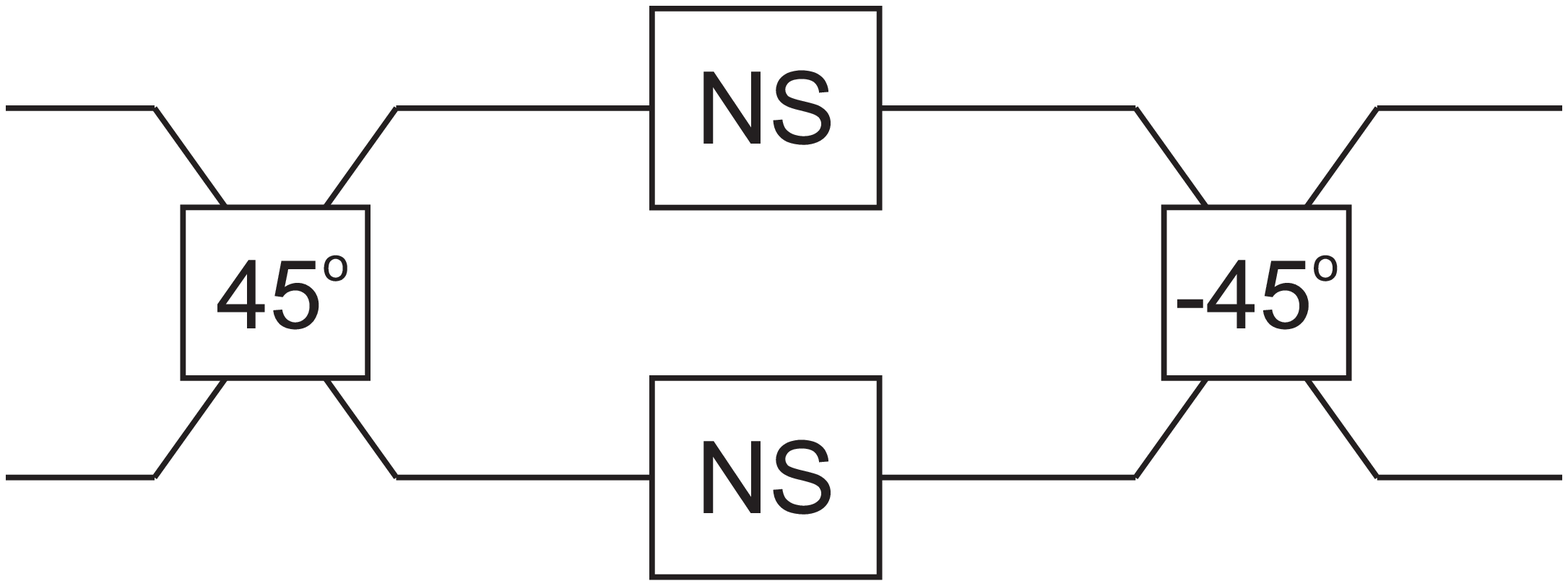,clip=}}
 \bigskip
 \caption{C-SIGN gate consisting of two beam splitters and two
non-linear phase shifters NS.}
 \label{C-SIGN}
\end{figure}

Using two beam splitters and two operation NS one can built
a conditional sign-flip gate (``C-SIGN'') as shown in Fig.\
\ref{C-SIGN}. This gate inverts the sign of the state vector of two
modes when there are exactly one photon in each of them. In the other
cases (with at most one photon altogether) the operation does
nothing:
$$
\begin{array}{lcrllcrl}
 |1\rangle |1\rangle &\to& -&|1\rangle |1\rangle, \quad &
 |1\rangle |0\rangle &\to&  &|1\rangle |0\rangle, \\
 |0\rangle |1\rangle &\to&  &|0\rangle |1\rangle, &
 |0\rangle |0\rangle &\to&  &|0\rangle |0\rangle.
\end{array}
$$
The unitary matrix representing the transformation of
creation operators on the beam splitter is supposed to be
$$
\left( \begin{array}{lr}
 \cos\theta & -\sin\theta \\ \sin\theta & \cos\theta
\end{array} \right).
$$
Particular angles $\theta$ are written inside the corresponding boxes
in Fig.\ \ref{C-SIGN}. Since this gate uses two NS operations its
probability of success is 1/16.

\begin{figure}
 \centerline{\psfig{width=\hsize,file=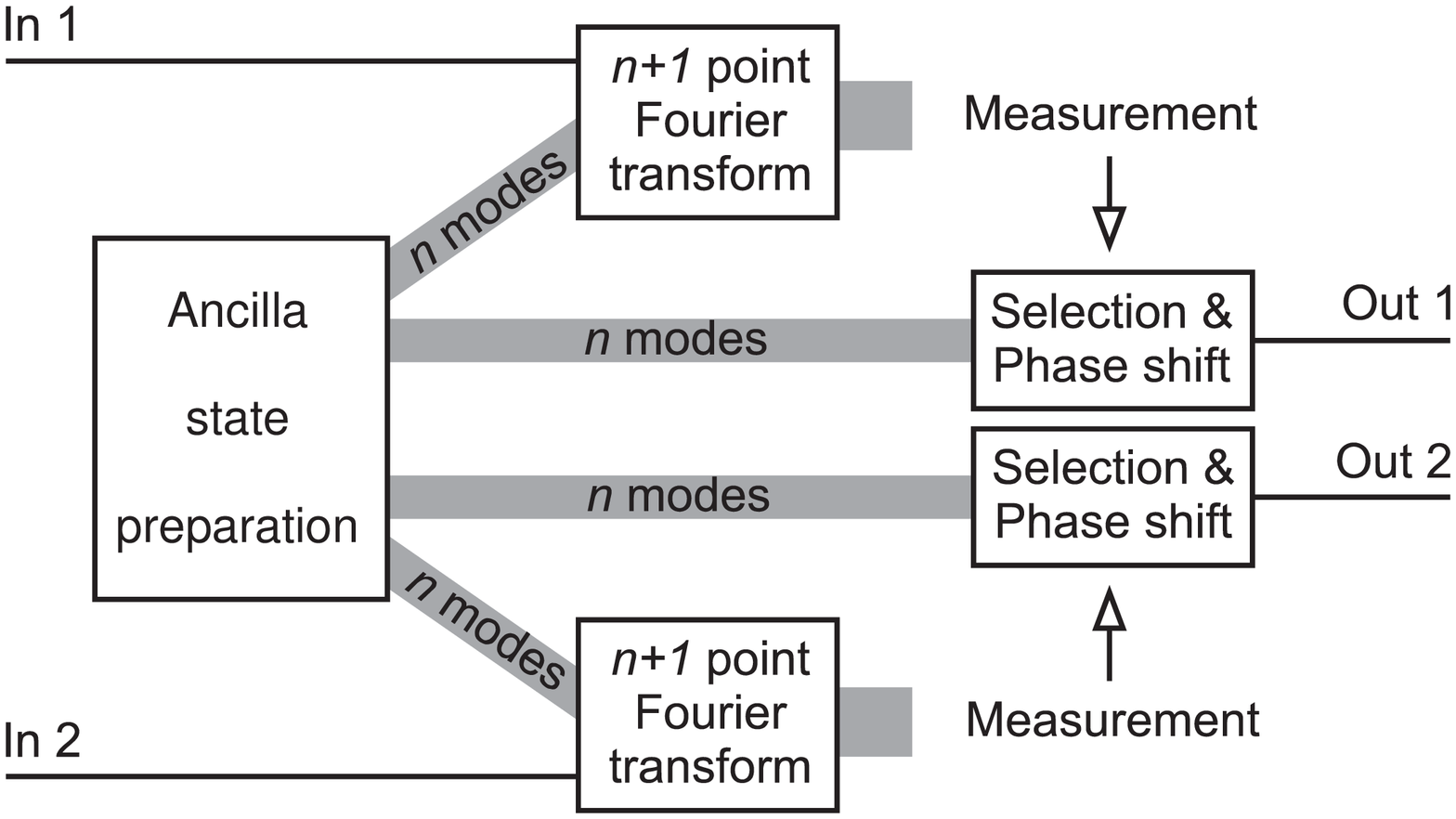,clip=}}
 \bigskip
 \caption{Schematic view of the setup for C-SIGN gate with the
probability of success $p=[n/(n+1)]^2$. Here $4n$ ancillary modes are
in the state given by Eq.\ (\ref{ancilla}).}
 \label{C-Sn}
\end{figure}

However, in the mentioned papers the way is proposed how to increase
the probability of the successful action of the C-SIGN arbitrarily
close to one. This way is based on a ``teleportation trick''. It
requires preparation of relatively complicated ancillary entangled
state of $2n$ photons in $4n$ modes:
\begin{eqnarray}
  |\phi_n\rangle = \sum_{j,k=0}^{n} (-1)^{(n-j)(n-k)}
  &&
  |1\rangle^j |0\rangle^{n-j} |0\rangle^j |1\rangle^{n-j}
  \nonumber  \\   \times~ &&
  |1\rangle^k |0\rangle^{n-k} |0\rangle^k |1\rangle^{n-k},
\label{ancilla}
\end{eqnarray}
where $|x\rangle^y = |x\rangle |x\rangle \dots |x\rangle$, $y$-times.
First $n$ modes are ``put together'' with one input mode
and all these $n+1$ modes are subjected to $n+1$ point Fourier
transform. Then the photon number measurement is performed on the
transformed modes and according to the result one of the other $n$
modes is chosen as an output of the gate and its phase is modified in
general. The same action is done with the next $2n$ modes and the
other input mode; see Fig.\ \ref{C-Sn}. It can be shown that the
total probability of the success of such a C-SIGN gate is
\begin{equation}
 p = \left( \frac{n}{n+1} \right)^{\!2}.
\label{prob}
\end{equation}
The Fourier transform, selection of modes, and phase shifts can be
implemented by linear optical elements in a deterministic way. The
state (\ref{ancilla}) can be prepared by means of NS operations, beam
splitters and phase shifters. The probability of successful
preparation can be rather low. But we should stress that this
concerns just the preparation of an ancilla. In principle, it can be
being prepared in advance and one can try many times.
For more details, including the estimation of the success
probability of the preparation procedure and the number of necessary
elements, see Refs.\ \cite{LOQC_nature,LOQC_qph}.

\begin{figure}
 \centerline{\psfig{width=\hsize,file=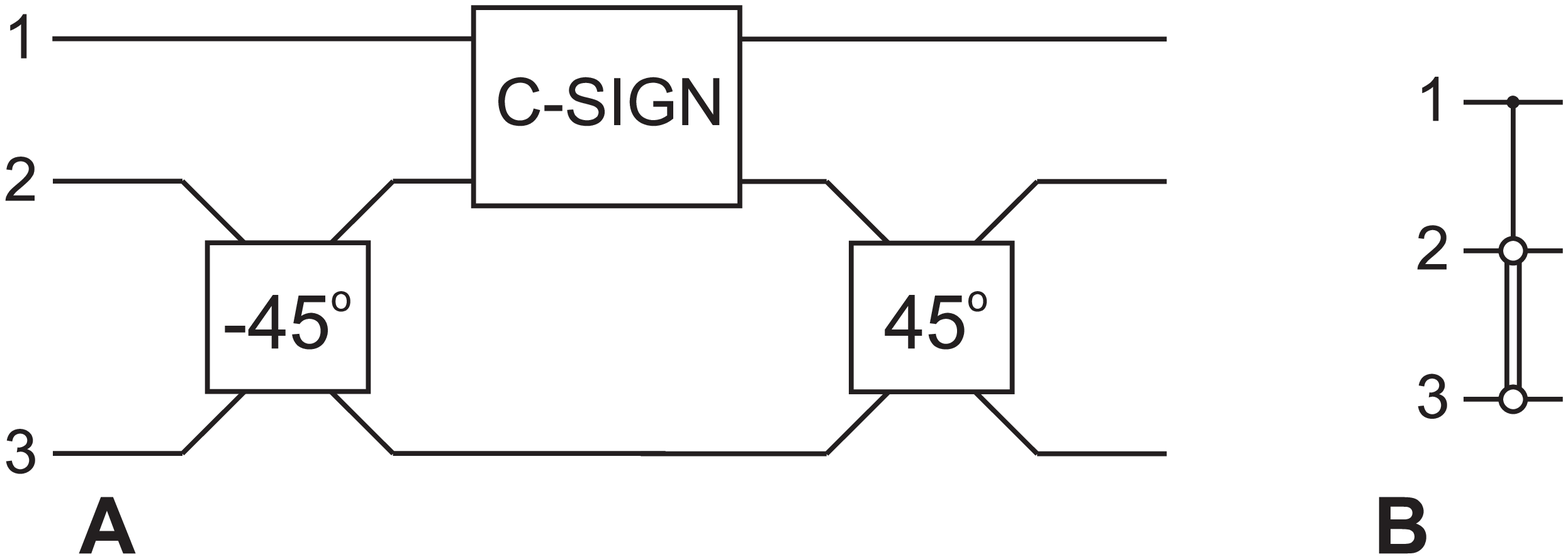,clip=}}
 \bigskip
 \caption{C-SWAP gate that swaps the modes 2 and 3 if there is a
photon in the mode 1. A: The scheme of the gate built from two
beam splitters and one C-SIGN gate. B: The notation we will use.}
\label{C-SWAP} \end{figure}

\begin{figure}
 \centerline{\psfig{width=0.6\hsize,file=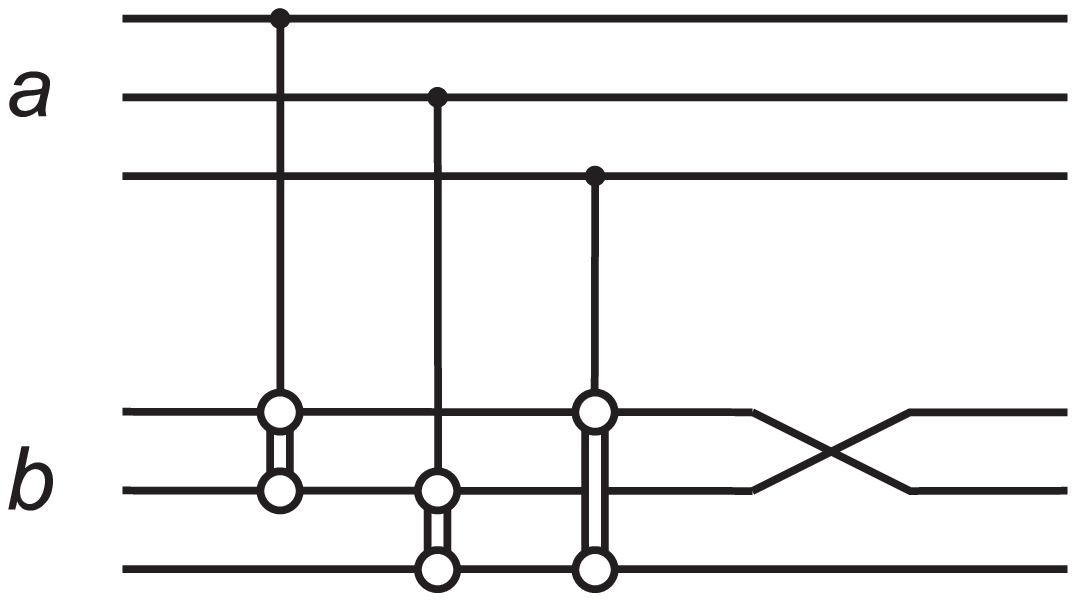,clip=}}
 \bigskip
 \caption{C-SHIFT gate. An example of the network for $d=3$ built-up
from C-SWAP gates. Here $a$ denotes the control qudit, $b$
the controlled one.}
 \label{C-SHIFT}
\end{figure}

Having a C-SIGN gate one can realize a gate ``C-SWAP'' that
conditionally swaps the two modes of radiation (provided there is at
most one photon altogether in them). Its scheme is in Fig.\
\ref{C-SWAP}. If there is one photon in the mode 1 the modes 2 and 3
are swapped. If no photon is present in the mode 1 they are not
changed. The probability of the success of the C-SWAP is the same as
for the C-SIGN. The gate C-SWAP is the key ingredient to construct a
logical operation acting on two qudits $x, y$ of arbitrary dimension
$d$ that we will call ``C-SHIFT'':
$$
x \to x, \quad y \to (y-x) \mbox{~mod~} d.
$$
For any $x$ this operation represents a cyclic
permutation or ``rotation'' of the values of the second qudit. In
total, $d-1$ rotations are necessary --- each of them corresponds to
one possible value of the first qudit except the value zero that
leads to identity transformation. Any such rotation can be
implemented by at most $d-1$ ``transpositions'', i.e., C-SWAPs. Thus
if the probability of success of the C-SWAP is $p$ then the total
probability of the success of the C-SHIFT is at least $p^{(d-1)^2}$.
In particular cases it can be even better, e.q., for $d=3$ the
probability of the success of the C-SHIFT is $p^{3}$. Corresponding
``network'' is in Fig.\ \ref{C-SHIFT}. If one uses C-SIGN gates
described above with the probability of success given by
Eq.\ (\ref{prob}) then the probability of the successful action of
the C-SHIFT gate is
\begin{equation}
 P \ge \left( \frac{n}{n-1} \right)^{\!2 (d-1)^2}.
\label{TotProb}
\end{equation}

\begin{figure}
 \centerline{\psfig{width=0.8\hsize,file=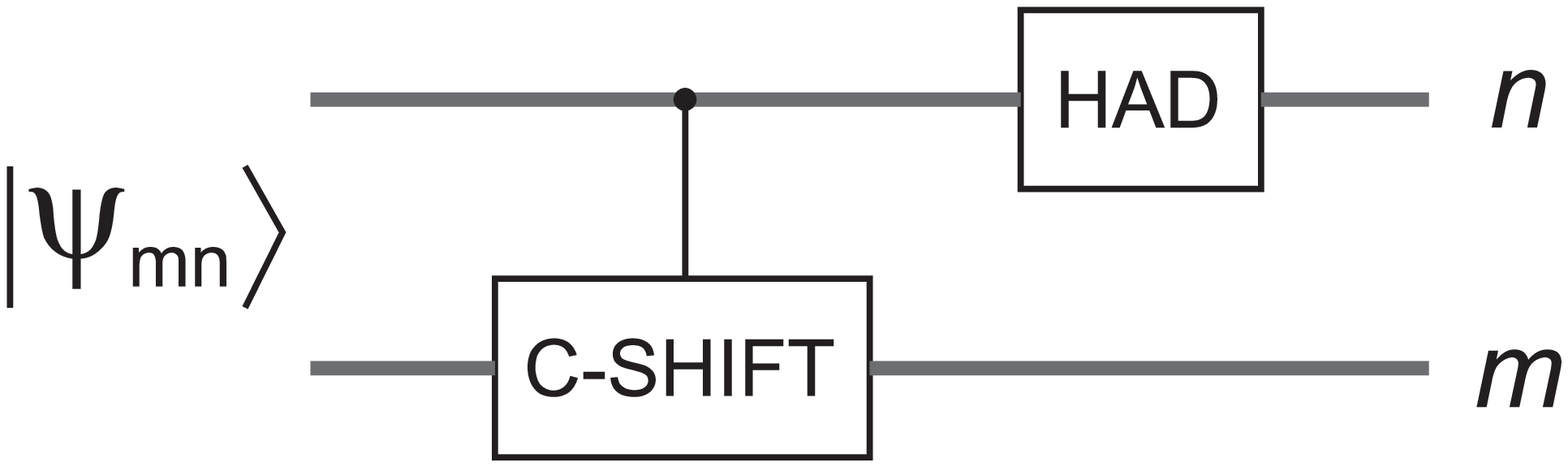,clip=}}
 \bigskip
 \caption{Scheme of the Bell-state analyzer for two qudits. It
consists of a C-SHIFT operation and a generalized Hadamard
transform.}
 \label{BA}
\end{figure}

Now we can build a Bell-state analyzer. To do it we need our C-SHIFT
gate,
\begin{equation}
 \mbox{C-SHIFT:~~} |x\rangle_1 |y\rangle_2 \to
                 |x\rangle_1 |(y-x) \mbox{~mod~} d\, \rangle_2 ,
 \label{shift}
\end{equation}
followed by a generalized Hadamard transform acting on the first
qudit,
\begin{equation}
 \mbox{HAD:~~} |x\rangle_1 \to \frac{1}{\sqrt{d}} \sum_{k=0}^{d-1}
             \exp \left[ -2 \pi i \frac{k x}{d} \right] |k\rangle_1,
 \label{had}
\end{equation}
where $|j\rangle_l$ represents the $j$-th ``logical'' state of the
$l$-th qudit. The complete setup is shown in Fig.\ \ref{BA}.
If the sequence of these two operations is applied on a Bell state
\begin{equation}
  |\psi_{mn}\rangle = \frac{1}{\sqrt{d}} \sum_{j=0}^{d-1}
  \exp \left[ 2 \pi i \frac{j n}{d} \right]
  |j\rangle_1 |(j+m) \mbox{~mod~} d\, \rangle_2
\label{Bell2}
\end{equation}
[see also Eq.\ (\ref{Bell})] then the output state
of the two qudits is $|n\rangle_1 |m\rangle_2$. Thus, if both the
qudits are realized as described above the input Bell state can be
determined by a simple photodetection.

The generalized Hadamard transform is the unitary transformation of
one qudit. As mentioned earlier any such operation can be realized
in a determistic way with passive linear optical elements (for our
implementation of qudits) --- simply by combination of beam splitters
and phase shifters. The C-SHIFT gate described above is a
non-deterministic gate. Its probability of successful action is given
by Eq.\ (\ref{TotProb}). Clearly, that value determines also the
probability of the successful discrimination of an unknown Bell
state. Increasing $n$ this probability can be made
arbitrarily close to one provided the dimension $d$ is fixed.

Let us note, that the described approach can be extended to the
discrimination of ``generalized Bell states'' of $N$ qudits, i.e.,
the following maximally entangled states that complete an orthonormal
basis in the Hilbert space of $N$ qudits:
\begin{eqnarray}
 |\psi_{k_1, k_2,\dots, k_N}\rangle &=& \frac{1}{\sqrt{d}}
   \sum_{j=0}^{d-1}   \exp \left[ 2 \pi i \frac{j k_1}{d} \right]
 \nonumber \\ & \times &
   |j\rangle_1 \bigotimes\limits_{i=2}^{N}  |(j+k_i) \mbox{~mod~} d\,
   \rangle_i,
 \label{N-Bell}
\end{eqnarray}
where $k_i=0,\dots,d-1$. In this case the C-SHIFT operation must be
applied $N-1$ times: between the first and the $N$-th qudit, the
first and the $(N-1)$-th qudit, etc. Finally between the first and
the second qudit. Then the generalized Hadamard transform must be
performed on the first qudit. If there was a generalized Bell state
(\ref{N-Bell}) in the input then the output state reads
$$
 |k_1\rangle_1 |k_2\rangle_2 \dots |k_N\rangle_N.
$$


\section{Conclusions}
\label{conc}

We have interpreted some recent results concerning unambiguous state
discrimination with linear optical elements from the point of view of
Bell-state measurement in case of two qudits with dimension higher
than two. This analysis leads to the conclusions that with no
auxiliary photons it is impossible to discriminate such Bell states
without errors and that it is impossible to discriminate such Bell
states with certainty in any way by the means of linear optics.

On the other hand, we have shown by an explicit construction that it
is possible, in principle, to build a linear optical Bell-state
analyzer capable to discriminate all the states of the Bell basis of
two (or even more) qudits with the probability of success arbitrarily
close to one. This device is based on the generalization of the idea
of linear optics quantum computation for qudits. The price for the
high success probability is a complicated setup and a large number
of required auxiliary photons in rather complex entangled states.

It seems that the key ingredient that is necessary for
the increase of the probability of successful discrimination is
the entanglement ``added'' through the ancilla. The methods of
``linear optics quantum computation'' enable us to prepare required
entangled states from ``separated'' photons, in principle. But the
probability of the success of such a preparation is very low. Besides
this, the preparation of single photons represents itself a serious
experimental problem as single-photon states are highly non-classical
ones. By the way, recently it was shown for all pure input states
and for the large class of mixed states that the beam splitter can
serve for the preparation of an entangled state on its output only if
the input state exhibits non-classical behavior \cite{kim}.


\section*{Acknowledgments}

The author appreciates discussions with John Calsamiglia and Norbert
L\"{u}tkenhaus. This research was supported under the European Union
project EQUIP (contract IST-1999-11053) and the project LN00A015 of
the Ministry of Education of the Czech Republic.



\end{document}